# Prior Knowledge based Advanced Persistent Threats Detection for IoT in a Realistic Benchmark

Yu Shen, Murat Simsek, Burak Kantarci, Hussein T. Mouftah, Mehran Bagheri, Petar Djukic

School of Electrical Engineering and Computer Science, University of Ottawa, Ottawa, ON, Canada

AI and Analytics Department of Ciena Corp., Ottawa, ON, Canada.

E-mails: ‡ {yshen041, murat.simsek, burak.kantarci, mouftah}@uottawa.ca, ‡ {mbagheri, pdjukic}@ciena.com

*Abstract*—The number of Internet of Things (IoT) devices being deployed into networks is growing at a phenomenal level, which makes IoT networks more vulnerable in the wireless medium. Advanced Persistent Threat (APT) is malicious to most of the network facilities and the available attack data for training the machine learning-based Intrusion Detection System (IDS) is limited when compared to the normal traffic. Therefore, it is quite challenging to enhance the detection performance in order to mitigate the influence of APT. Therefore, Prior Knowledge Input (PKI) models are proposed and tested using the SCVIC-APT-2021 dataset. To obtain prior knowledge, the proposed PKI model pre-classifies the original dataset with unsupervised clustering method. Then, the obtained prior knowledge is incorporated into the supervised model to decrease training complexity and assist the supervised model in determining the optimal mapping between the raw data and true labels. The experimental findings indicate that the PKI model outperforms the supervised baseline, with the best macro average F1-score of 81.37%, which is 10.47% higher than the baseline.

## I. INTRODUCTION

Unlike common network attacks, which have a limited attack period, the Advanced Persistent Threat (APT) has a prolonged attack time, disguised among normal traffic patterns so as to pose serious data leakage in a network [1]. There are six principal stages in the APT and different attack stages can be transformed into each other. The six principal stages of APT are demonstrated in Fig. 1. Due to the adaptability of the attack strategy, it is challenging to acquire appropriate data for APT detection. The term APT was initially coined in 2006 [2], and it is implemented to intrude networks of military organizations and stole non-public information originally. Government departments have been targets of APT due to the invaluable and nonpublic information [3]. The leakage of the information and data causes unpredictable losses to relevant departments [4]. Aside from the internal networks of enterprises, Internet of Things (IoT) environments such as smart grid and Industry Control System (ICS) are also vulnerable to APT attacks due to the prevalence of devices that are equipped with smart decision support systems [5]. In a large-scale IoT environment, the vulnerabilities of the underlying the communication protocols utilized by smart sensors and IP cameras could serve as initial compromise sites for APT attackers.

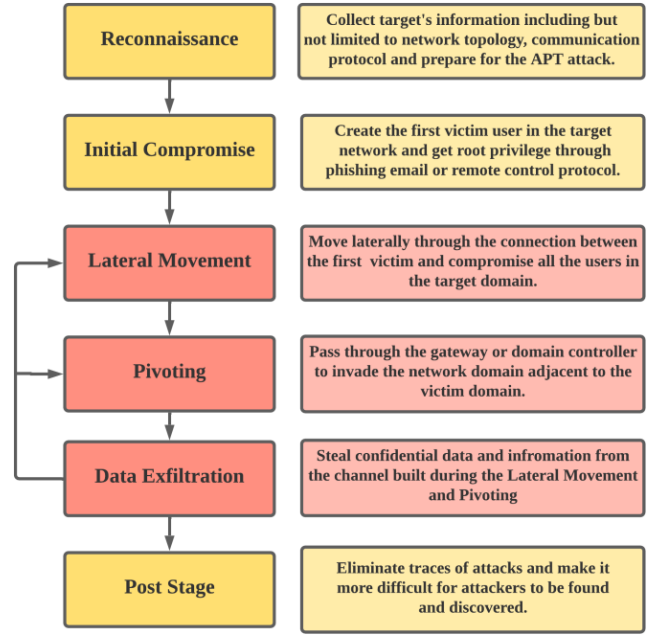

Fig. 1: APT attack stages. Yellow Blocks: initial and final stages. Red Blocks: attack stages.

The defence strategies against APT attacks are primarily divided into monitoring, mitigation and detection methods [6]. The monitoring methods entail detecting APT by analyzing the hardware information and event traces generated by the devices. PCs and anti-virus software are the primary consumers of the monitoring methods. The mitigation methods are described as taking reactions to lessen the impact of APT. Two instances of the mitigation methods are related to attack graphics and fake network files [6]. It is worth noting that both the monitoring and mitigation methods have inherent downsides. When there are a significant number of network users or when network traffic is excessively complicated, the monitoring methods lead to a relatively high false-positive rate, which complicates real-time monitoring. Since the mitigation methods are incapable of actively detecting attacks, they are used to mitigate losses only after part of the network is compromised.

Due to the shortcomings of the monitoring methods and mitigation methods that are illustrated above, the machine learning detection methods are proposed [7]. The trained model is capable of monitoring the coming network packets in real time with high accuracy in order to raise an early alert prior to data exfiltration. Because of the APT attack data deficiency, conventional machine learning strategies are not adequate to efficiently detect the APT attack. The motivation of this study is to incorporate more knowledge in detection systems to improve the overall performance of defense strategies. In light of this, the contributions of this paper are listed below:

- A knowledge-based neural network, namely Prior Knowledge Input (PKI) is introduced into the detection model to combine additional knowledge provided by unsupervised clustering algorithms with existing features to detect the APT. Since PKI is originally developed to reduce the computational complexity of engineering problems, it is applied to this problem to boost the detection performance under limited available APT attack data.
- SCVIC-APT-2021 dataset [8] [1] that was originally generated in a laboratory setting covers five APT stages: Initial Compromise on a node that represents the services used by smart sensors and IP cameras (i.e., an IoT context), Pivoting, Lateral Movement, Reconnaissance, and Data Exfiltration. The performance evaluation criteria is chosen as F1-score to take into account false predictions during the testing phase. The highest macro average F1-score under the SCVIC-APT-2021 dataset is 81.37%, which is 10.47% higher than the supervised baseline performance.

In the remainder of this paper, Section II demonstrates related work on APT attack detection using machine learning. Section III describes the PKI and Progressive PKI models' workflow. Additionally, the evaluation criteria for the experimental outcomes are introduced. The numerical results and conclusions are given out in Section IV and Section V respectively.

## II. RELATED WORK

### A. Machine Learning for APT Detection

Machine learning algorithms are more adaptable, since they can evaluate a single network flow or host event in real time [9]. T. Bodström et al. propose a deep learning stack for detecting APT attacks [10]. This approach detects APT layer-by-layer by first operating in serial mode and then in parallel mode. When the model detects an APT attack, it sends the data to the attack database for storage and analysis.

Due to the malicious payload transmission happening prior to data exfiltration, the APT can be identified indirectly by detecting the presence of malicious payload. Lu et al. propose using time transform features to distinguish normal network traffic and network traffic with malicious payloads [11]. They capture normal traffic passing through university gateways and merge it with APT traffic before feeding them to machine learning algorithms. The results indicate that temporal transform features have the potential to significantly improve detection performance.

---
[1]The SCVIC-APT-2021 dataset in [8] is publicly available at https://ieee-dataport.org/documents/scvic-apt-2021.

Matsuda et al. propose a method for detecting APT with host events [12]. Attackers can utilize Active Directory (AD), a centralized administration system for Windows computers, to appear as legal administrator account users, facilitating data leakage. This type of APT, on the other hand, leaves user login and logout records on the host events system, which can be used to detect intrusions.

### B. ML- and Knowledge-Based models for IoT security

In a recent survey, we have summarized the APT attack types and machine learning-based APT detection solutions in IoT environments [5]. The study was pursued by using a specific threat analysis model, namely Process for Attack Simulation and Threat Analysis (PASTA), whichbuilds on seven layers to enable analysis of the impact of APT in a top-down manner. Generally, both supervised and unsupervised methods are used to safeguard IoT systems against APT attacks. Indeed, with the advent of improved computational power with the availability of graphical processing units, deep learning-based solutions such as Deep Neural Network (DNN) and Convolutional Neural Network (CNN) have considerably advanced the performance of predictive solutions [13].

The knowledge-based models are used to shorten the training time of neural networks and simplify the complexity of the link between input and output [14]. The implementation of the knowledge-based models boosts the performance of the baseline model and reduces the training complexity. This study is inspired by the following knowledge-based solutions in other areas.

In a distinct IoT setting, namely mobile crowdsensing (MCS), Simsek et al. propose a method for detecting fake sensing tasks uploaded to the MCS servers that combines a Deep Prior Knowledge Input (Deep-PKI) with a Self-Organizing Feature Map. As a result, the Deep-PKI model helps the MCS server filter out identified fake tasks and does not allocate them to participants. The proposed method outperforms the baseline accuracy of deep neural networks [15].

## III. METHODOLOGY

### A. Prior Knowledge Input

Machine learning aims to replicate the principle of human cognition when it comes to new objects by building algorithm models for regression or classification. After the model is trained by Prior Knowledge Input (PKI) [16], fewer features can be utilized to obtain the same or better classification results. Therefore, the training complexity is reduced and the training efficiency is increased. The PKI model is extensively utilized to embed the prior knowledge besides the existing features [17] in an efficient neural network modelling process.

The training and testing phases of the PKI model are represented in Fig. 2. To begin with, the unsupervised model is trained for prior knowledge generation The number of clusters is the most crucial parameter for the unsupervised clustering algorithm and it directly affects the performance of the unsupervised model. Another important aspect is the validation set that is separated from the training set in order to determine the appropriate number of clusters for the unsupervised model. Following upon training of the unsupervised model, the supervised model is trained with both existing features and prior knowledge generated by the unsupervised model. The prior knowledge provides regional knowledge based on similarity of each sample so the supervised model with prior knowledge can establish more accurate input-output relationship than those without prior knowledge during the training phase.

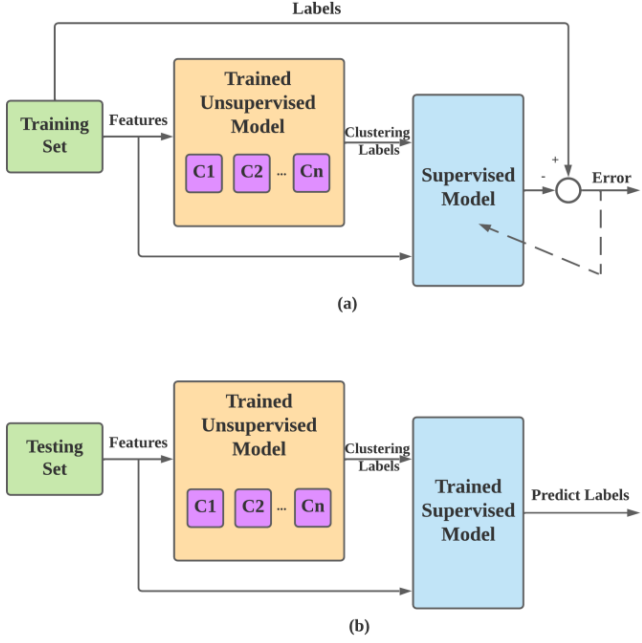
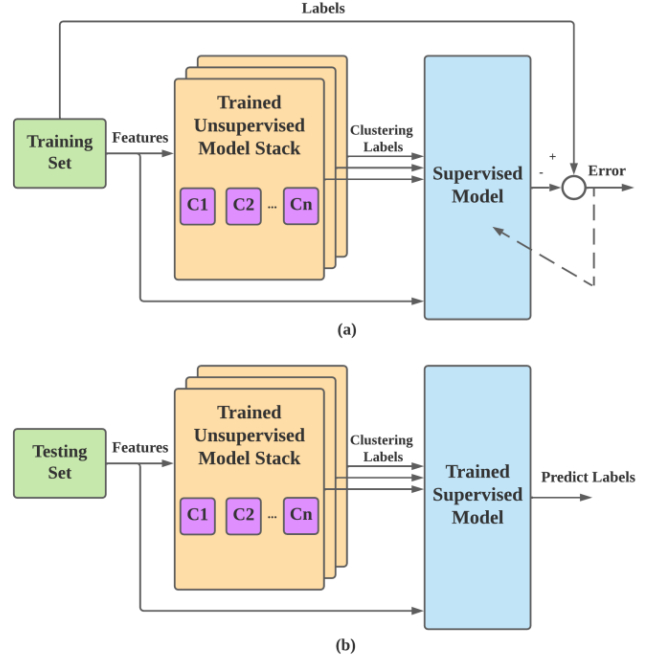

Fig. 2: (a) The training phase of the PKI model, (b) The testing phase of the PKI model

Fig. 3: (a) The training phase of the Progressive PKI model, (b) The testing phase of the Progressive PKI model

The testing phase of PKI is substantially identical to training. The only difference is that the testing phase does not require to assess the error and to update the parameters of the supervised model. During the testing phase, the features of all data points are fed into the unsupervised model in order to obtain the prior knowledge. Then, prior knowledge and original features are fed into the supervised model for prediction.

*B. Progressive PKI*

The Progressive PKI model is dependent on the PKI model and utilizes an unsupervised model stack to obtain prior knowledge. In the PKI model, clustering labels generated by a single unsupervised model are used to reduce training complexity. The SCVIC-APT-2021 dataset contains more than 80 features. Due to the prior knowledge obtained by the PKI model being limited to a single dimension, it is difficult to improve the classification performance of the model in comparison to the huge number of original features. Additionally, if the clustering of the unsupervised model is unable to adequately reflect the true distribution of data points, the performance of the model degrades. As a result, Progressive PKI is developed to solve the prior knowledge dimensionality limitations of the PKI model.

The training and testing phases of the Progressive PKI model are presented in Fig. 3. Prior to the training phase, the unsupervised model stack is trained. The unsupervised model stack is composed of multiple unsupervised models that are totally independent of each other. Each unsupervised model obtains prior knowledge by clustering the original features of the dataset. The crucial step in this approach is to determine the size of the unsupervised model stack. The size indicates the number of unsupervised models included in the stack. In order to obtain the optimal size, a validation set is separated from the training set and used for tuning the stack size. After training the unsupervised model stack, all prior knowledge from the unsupervised stack and original features from the dataset are fed into the supervised model for traffic classification training.

During the testing phase, the original features of the dataset and the prior knowledge generated from the unsupervised model stack are input into the trained supervised model to predict whether the network flow is malicious.

*C. Evaluation Criteria*

Normal Traffic occupies 98.36% of the dataset in order to imitate the real network environment, resulting in the imbalance problem. The average evaluation measures, such as precision and recall, do not capture the true performance of the model. For instance, if the model classifies all data points as Normal Traffic, it can still produce excellent average precision and recall, but the attack detection accuracy is 0.

The Macro Average F1-score is used as the criterion for evaluation. As formulated in (1), it is defined as the average of F1 scores of all classes where F1 score denotes the harmonic mean of precision and recall parameters, i.e., $F1\ score = (2 \times Precision \times Recall)/(Precision + Recall)$. The macro average of F1-scores of all classes can be used to solve the problem of evaluation criteria caused by data imbalance.

$$Macro\ Avg\ F1\ score = \frac{\sum_{i}^{m} F1\ score^{(i)}}{m} \quad (1)$$

## IV. NUMERICAL RESULTS

In this study, Scikit-Learn ML package is used on an i7-6700HQ 2.60 GHz CPU with 8GB RAM computer running 64-bit Windows10. The SCVIC-APT-2021 dataset and experimental results are discussed below.

### A. SCVIC-APT-2021 Dataset

The laboratory-created dataset, SCVIC-APT-2021[2], which is utilized in the following experiment primarily covers three APT attacks: Lateral Movement, Pivoting, Data Exfiltration, and normal data packets. While the data packets generated by hackers during the attacks on the network domain entry host is named as Initial Compromise, the investigation phase within the hacked domain is named as Reconnaissance. A connected IP camera provides IoT communication packets within the network domain via the host of entrance. As a result, the dataset has six distinct types. The class distributions for the SCVIC-APT-2021 dataset are shown in Table I.

TABLE I: SCVIC-APT-2021 Dataset Class Distribution

| Classes | Training Set | Testing Set | Total |
|---|---|---|---|
| Data Exfiltration (DE) | 527 | 74 | 601 |
| Initial Compromise (IC) | 73 | 77 | 150 |
| Lateral Movement (LM) | 729 | 142 | 871 |
| Normal Traffic (NT) | 254836 | 55583 | 310419 |
| Pivoting (P) | 2122 | 360 | 2482 |
| Reconnaissance (R) | 833 | 251 | 1084 |

**Preprocessing**: The comma-separated values (CSV) file generated by CICFlowmeter contains the node hardware information, such as IP address and port number. These features are node-specific. For example, if a hacker takes control of a particular port, the data packets received from that port are likely to be harmful to the computer system. If these hardware features are fed into the model during the training phase, the model would judge that all the packets from that specific port are malicious, not according to the network flow features. This scenario disturbs the learning from network flow features and reduces the scalability of the machine learning model. To enable the model to be deployed successfully in a variety of network contexts, the hardware information associated with each data point is erased. For the machine learning training process, the preprocessed dataset contains a total of 76 features.

**Supervised Baseline**: Three supervised learning algorithms, Decision Tree (DT), Random Forest (RF) and XGBoost (XGB), are applied to the preprocessed dataset to form the baseline.

[2]The SCVIC-APT2021 dataset will be made publicly available on Github. At the moment, reviewers can access it via the following link: https://tinyurl.com/4ba9vw2y

TABLE II: The Baseline 1 macro average F1-scores

| Class | DT | RF | XGB |
|---|---|---|---|
| DE | 0.288 | 0.301 | 0.268 |
| IC | 0.800 | 0.829 | 0.861 |
| LM | 0.785 | 0.782 | 0.881 |
| NT | 0.999 | 1.000 | 1.000 |
| P | 0.733 | 0.851 | 0.757 |
| R | 0.403 | 0.749 | 0.490 |
| W Avg | 0.993 | 0.996 | 0.994 |
| M Avg | **0.668** | **0.752** | **0.709** |

TABLE III: The Baseline 2 macro average F1-scores for SCVIC-APT-2021 (Feature Selection Results)

| Supervised Methods | Feature Selection | Optimal Number of Features | Macro Avg F1-score |
|---|---|---|---|
| RF | **Chi2** | **51** | **0.8034** |
|  | ANOVA | 50 | 0.7738 |
|  | Mutual Info | 12 | 0.7883 |
| XGB | **Chi2** | **49** | **0.8092** |
|  | ANOVA | 19 | 0.7616 |
|  | Mutual Info | 30 | 0.7310 |

When the preprocessed dataset is fed into DT, RF and XGB, the macro average F1-scores of the supervised baselines are 0.668, 0.752, and 0.709, respectively. The supervised model baselines are called Baseline 1, which serves as a reference result for subsequent results. Table II illustrates the numerical results of Baseline 1. W Avg and M Avg represent Weighted Average and Macro Average, respectively. RF and XGB are two ensemble methods based on DT and achieve better performance than DT on the preprocessed dataset according to the comparison of the results. Therefore, RF and XGB are reserved in the subsequent experiment and DT is dropped because of the low performance.

**Feature Selection**: Following the acquisition of Baseline 1, filter-based feature selection approaches such as Chi2 [18], ANOVA [19], and Mutual Information [20] are used to help simplify the dataset and eliminate redundant features. For each feature selection method, we go through all possible number of features and select the best one with the highest F1 score on the validation set. SCVIC-APT-2021 contains 76 network features after preprocessing. Chi2, ANOVA, and Mutual Information are used to choose from 1 to 76 features from the dataset separately. The optimal number of features for the SCVIC-APT-2021 are demonstrated in Table III. The bold rows in the table represent the ideal combinations of the supervised model, the feature selection method, and the optimal number of features. It is found that when the Chi2 algorithm is used, the selected number of features for RF is 51. The highest macro average F1-score is 80.34%. Similarly, XGB reaches the best performance when 49 features are selected for Chi2. The best macro average F1-score is 80.92%. The selected features are transmitted to the PKI model in order to boost the performance of network attack detection. These optimal combinations are used as Baseline 2 for the following PKI model.

**Prior Knowledge Input**: The PKI experiment is dependent on the feature selection results. Unsupervised models are used to cluster the selected features produced from feature selection and then supervised models are used to predict the final classes using the clustering labels and selected features. It is critical for unsupervised models to pick the appropriate number of clusters. The validation set separated from the training set is utilized to determine it. In this experiment, k-means clustering (KMeans) and Gaussian Mixture Model (GMM) are two candidates for the unsupervised model. RF and XGB are candidates for the supervised model. Therefore, a total of four combinations are tested. The number of clusters varies from 2 to 20 during the experiment. The PKI model results on SCVIC-APT-2021 are illustrated in Table V. The bold rows in the tables are the highest performance that the PKI model can reach. The best macro average F1-score is 81.03% when the GMM (number of clusters = 5) is combined with RF. The confusion matrix for this combination is shown in Fig. 5. When KMeans and RF are combined, and the number of clusters of KMeans is 17, the highest macro average F1-score of this combination is 80.33%. However, for XGB, no matter which unsupervised model is used for prior knowledge clustering, the results obtained are lower than the results of Baseline 2. Therefore, the optimal number of clusters for GMM is 1 when combined with XGB and the detection performance equals Baseline 2.

TABLE IV: Parameters tuning for GMM and XGB combination

| Algorithms | Parameter | Candidate Set |
|---|---|---|
| GMM | Covariance Type | "spherical", "diag", "full", "tied" |
| XGB | Number of Estimator | Range: (10, 200), Step Size: 10 |
| | Learning Rate | 0.01, 0.05, 0.1, 0.2, 0.3 |

**Progressive PKI**: Fig. 3 indicates that multiple columns of clustering labels obtained via the unsupervised model stack are integrated into the original dataset as new features to aid in detecting APT. The optimal results are illustrated in Table VI after traversing the size of the unsupervised model stack from 1 to 20. For XGB, it is discovered that the Progressive PKI model performs better than the PKI model. The combination of GMM and XGB produces the best macro average F1-score, 81.27%. Furthermore, Fig. 4 compares the detection performance of Baseline 1, Baseline 2, PKI model and Progressive PKI model. An extraordinary APT detection improvement can be found when the PKI and Progressive PKI models are applied.

**Parameter Tuning**: According to previous findings, the combined GMM and XGB Progressive PKI model achieve the highest macro average F1-score. The parameters are tuned for this combination to maximize APT detection performance. We define a set within an appropriate range for each parameter and then use grid search to identify the optimal parameter combination. The adjustment range and step length for each parameter can be found in Table IV.

During the grid search, 0.2 is selected as the optimal learning rate for XGB as the learning rates of 0.3 or smaller than 0.2, the performance of Progressive PKI-based XGB leads to performance degradation. Regarding the number of estimators of XGB, the default value of 100 is chosen as the optimal for the proposed model during the grid search. Therefore, when the covariance type of GMM is "full", the number of estimators, and the learning rate of XGB are 100 and 0.2, respectively, the best macro average F1-score is 81.37%. The classification confusion matrix is depicted in Fig. 6. It is found that several more data points in Data Exfiltration, Normal Traffic and Reconnaissance classes are classified correctly when compared to the RF PKI model in Fig. 5. Although the performance of the XGB Progressive PKI model on the Lateral Movement and Pivoting is degraded a bit, it is still the highest macro average F1-score that the proposed methods obtained in this experiment.

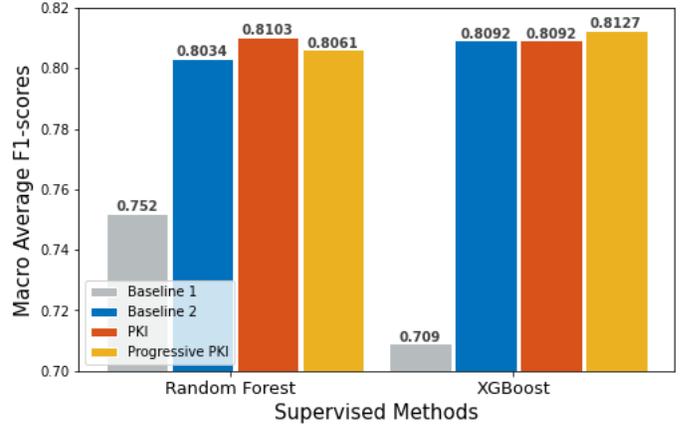

Fig. 4: Macro average F1-score comparison of all methods

Fig. 5: Confusion Matrix for Random Forest PKI

## V. CONCLUSION

As the APT attack initiators can modify their strategies based on the real-time environment of the victim network, there is an urgent demand for accurate APT attack detection on IoT environments. This study proposes a Prior Knowledge Input (PKI)-based APT detection in an IoT context. The PKI and a Progressive PKI model utilize unsupervised learning to cluster features and reduce the complexity of training so that the model performs better with fewer features. The experimental results

TABLE V: The PKI model results for SCVIC-APT-2021 dataset

| Supervised Method | Feature Selection | Number of Features | Unsupervised Methods | Optimal Number of Clusters | Macro Avg F1-score |
|---|---|---|---|---|---|
| RF | Chi2 | 51 | KMeans | 17 | 0.8033 |
|  |  |  | **GMM** | **5** | **0.8103** |
| XGB | Chi2 | 49 | KMeans | 1 | 0.8092 |
|  |  |  | GMM | 1 | 0.8092 |

TABLE VI: The Progressive PKI model results for SCVIC-APT-2021 Dataset

| Supervised Method | Feature Selection | Number of Features | Unsupervised Method | New Added Columns of Prior Knowledge | Macro Avg F1-score |
|---|---|---|---|---|---|
| RF | Chi2 | 51 | KMeans | 3 | 0.8046 |
|  |  |  | GMM | 2 | 0.8061 |
| XGB | Chi2 | 49 | KMeans | 9 | **0.8127** |
|  |  |  | GMM | 1 | 0.8088 |

Fig. 6: Confusion Matrix for XGBoost Progressive PKI

Confusion matrix values:
- DataExfiltration: 35, 0, 0, 2, 16, 21
- InitialCompromise: 0, 62, 0, 8, 4, 3
- LateralMovement: 0, 0, 112, 14, 3, 13
- NormalTraffic: 0, 0, 0, 55577, 5, 1
- Pivoting: 4, 0, 2, 12, 340, 2
- Reconnaissance: 23, 1, 6, 8, 41, 172

Columns: DE, IC, LM, NT, P, R

also illustrate that the introduced PKI and Progressive PKI models significantly improve the APT detection performance under a dataset generated in an IoT context. The PKI model increases the macro average F1-score of RF from 75.20% to 81.03%. For the Progressive PKI model, it improves the macro average F1-score of XGB from 70.90% to 81.37%.

ACKNOWLEDGEMENT

This work is supported in part by the Ontario Centre for Innovation (OCI) under ENCQOR 5G Project #31993.